\documentclass[pre,twocolumn,showpacs,preprintnumbers,amsmath,amssymb]{revtex4}

\usepackage{graphicx}

\begin{document}

\title{Fluctuations of grains inside a discharging two-dimensional silo}

\author{Angel Garcimart\'{\i}n}
\email{angel@fisica.unav.es}
\author{Iker Zuriguel}
\author{Alvaro Janda}
\author{Diego Maza}

\affiliation{Departamento de F\'{\i}sica, Facultad de Ciencias,
Universidad de Navarra, 31080 Pamplona, Spain.}

\date{\today}

\begin{abstract}
We present experimental data corresponding to a two dimensional dense granular flow, namely, the gravity-driven discharge of grains from a small opening in a silo. We study the microscopic velocity field with the help of particle tracking techniques. From these data, the velocity profiles can be obtained and the validity of some long-standing approaches can be assessed. Moreover, the fluctuations of the velocities are taken into consideration in order to characterize the features of the advective motion (due to the gravity force) and the diffusive motion, which shows nontrivial behaviour.
\end{abstract}

\pacs{45.70.-n}

\maketitle

\section{Introduction}
\label{sect:intro}

The process of silo discharge can be considered a canonical example of the complexity displayed by granular flows \cite{Duran}. Typically, when an orifice is opened at the bottom of a container filled with grains, the particles will exit through it and the outflow gives rise to a velocity field inside the silo. For big orifices the flow will quickly homogenize, while for small orifices it will develop intermittencies \cite{Rim}. A clear-cut definition allowing to qualify what \emph{big} or \emph{small} really means is unavailable, but it is normally accepted that an orifice diameter of about five times the typical size of the beads could be taken as a boundary between both situations. Recent systematic works suggest that even larger orifices are needed to avoid jamming events and intermittencies \cite{To2,Zuriguel,Janda1}. Regardless of these two regimes, the prediction of the flow rate at the outlet remains an open problem, at least in the sense that a derivation of a quantitative expression from fundamental principles has not been worked out yet. Simple dimensional analysis \cite{Beverloo} provides a relationship between the orifice size and the mass flow, but a theoretical foundation is lacking to support it. One of the reasons for the absence of this theoretical framework is that the flow near the outlet is accelerated due to the breakdown of the balance between gravity and dissipation. Nevertheless, far from the exit orifice (far meaning a few orifice diameters above it), the stationary velocity field is well defined if the outlet size is big enough. This suggests that the velocity profile could be described as a function of the stress field. However, an equation relating both variables has proven elusive. Perhaps the most important reason for that is the absence of a suitable spatial scale where to average variables like the density or the velocity in order to provide the likeness of an hydrodynamical description.

Two utterly different approaches have been employed to predict the velocity profile inside a silo. The first one --profusely used in soil mechanics-- starts from the consideration of the granular material as a continuous deformable media, and leads to concepts like yield stress and plastic potentials \cite{Nedderman}. Despite its success in generally describing many situations, some concerns remain about its applicability whenever discontinuities in the stress and velocity fields appear, which are not unusual. The second approach assumes from the very beginning the discrete nature of the material. The collective movement is taken as the sum of individual particles. It is easier to consider the diffusion of ``voids'' instead of particles; these voids enter in the silo through the outlet orifice and move upwards \cite{litw}. This notion is certainly suggestive and, once developed, predicts qualitatively well the velocity profile \cite{mullins}. Unfortunately, it fails to match quantitatively the experimental measurements of the velocity.

Nedermann and T\"{u}z\"{u}n \cite{kinematic} proposed the simple argument that the horizontal velocity component $u$ should be proportional to the gradient of the vertical velocity component $v$ along the horizontal direction, \emph{i. e.}
\begin{equation}
\label{eq:uv}
u=-B\frac{\partial v}{\partial x}
\end{equation}
where $B$ is a constant with length dimensions.  Hence, assuming that the flowing material is incompressible, it is easy to obtain a parabolic PDE for the spatial dependence of $v$:

\begin{equation}
\frac{\partial v}{\partial y}=-B\frac{{\partial}^2 v}{{\partial}^2 x}
\end{equation}

As this expression is isomorph to the diffusion equation, the constant $B$ is usually referred to as a ``diffusion length'', although such an statement is not rigorously true, given that $B$ was introduced just as an \emph{ad hoc} constant in Eq.~\ref{eq:uv}. From these premises, and assuming that the size of the orifice is negligible compared to the silo dimensions, the velocity profile in the vertical direction is:

\begin{equation}
\label{eq:vy}
v = - \frac{Q}{\sqrt{4 \pi B y}} \exp{\left( -\frac{x^2}{4 B y} \right)}
\end{equation}

where $Q$ is the volumetric flow rate. Notably, this expression has been successfully used to fit some experimental observations with an acceptable accuracy. The main shortcoming of this formula when confronted to the measurements is the wide discrepancy among the values of $B$ reported by different authors \cite{Choi1}. In some cases, a dependence of $B$ with the vertical coordinate has also been noted \cite{medina}.

Efforts have been recently made in order to find some common ground among all those lines of reasoning. As the concept of diffusive motion of individual particles or the holes left by them is broadly accepted in this context, some authors have generalized this notion by proposing the collective motion of groups of particles \cite{spot}. These authors put forward the existence of ``spots'', which are zones where the vacant space left by the absence of a particle is shared by a group of grains, causing a local decrease of the packing fraction. Such spots would diffuse upwards. In addition, they introduce a new length scale in the problem which allows to solve the discrepancies between the values of $B$ measured in the experiments and those predicted by the old diffusive approach. If diffusion is mediated by spots, the displacements of the particles are not necessarily of about their own size; moreover, the scale of the spot displacement can be \emph{quantitatively} related to the constant $B$ introduced by kinematic arguments.

No matter how valuable this idea is, the microscopic foundations of the  velocity profile and the mass flow rate remain unclear. Diffusive models are certainly inspiring but do not describe correctly important features of the flow, such as the fluctuations of the velocity field or the mean square displacement of the particles at short scales. These magnitudes have been considered in a series of recent papers
reporting anomalous PDF for the velocity distribution around its mean values \cite{choi,moka,index}. Also, the mean square displacements reported suggest that at short times the particle rearrangements are not ballistic and therefore the global behavior of the material can not be regarded as a classical hydrodynamic flow. Although the accuracy of some of the cited references represent a qualitative leap compared to earlier works, the experimental conditions used to obtain them prevent a direct comparison with theoretical models. For instance, it is sometimes argued that polydisperse grains are used to avoid crystallization effects; nevertheless, the disorder introduced by this practice can not be easily taken into account by any existing microscopic model. Besides, many of these experiments are performed in a quasi two dimensional configuration that could display in fact three dimensional effects, and the influence of this dimension on the flow properties is unclear. Finally, let us mention that recent approaches suggest the possibility of understanding the particle dynamics by means of elastoplastic hybrid ideas \cite{plastico} or a high density kinematic theory \cite{kumaran}.

In this work we attempt to avoid the mentioned troubles by performing our experiment in a single layer of particles. Although a strict 2D configuration is experimentally unaccessible, we designed a silo where the dynamics is driven only by the rearrangements of the particles within a single layer, and the effect of the front and rear walls just reduces to an extra dissipative term. We also use monodisperse particles, which indeed tend to crystallize, but as we will show in the following, both the microscopic and the macroscopic behavior of the granular flow are compatible with the above mentioned descriptions. We restrict ourselves to an orifice that is big enough to guarantee that the jamming probability is negligible.

The paper is organized as follows. We will first describe the experimental setup. Then we will show and comment the measurements corresponding to the velocity profiles. A section follows where we will focus on the velocities of individual particles. Then we will describe the fluctuations, and finally we will gather some conclusions.

\section{Experimental procedures and methods}
\label{sec:setup}

\begin{figure}
\includegraphics[width=0.9\columnwidth]{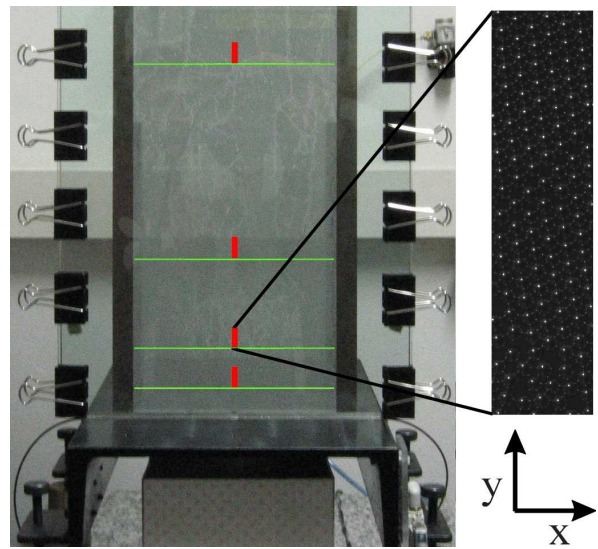}
\caption{(Color online) Sketch of the silo. The photograph covers approximately the half bottom of the silo. The two vertical strips at the sides are steel sheets that contain the beads and set the separation between the front and rear glass panes. The rectangular red patches show the four zones where images where recorded, which are described in the text. A typical image from one of the films is shown at the side. Horizontal green lines show the places where the measurement of the velocity profile was performed (see text). The choice of coordinates is also shown. The origin of the coordinates coincides with the center of the exit orifice.}
\label{fig:sistexp}
\end{figure}

As the experiment aim is the study of the movement of individual particles, we have built a two dimensional silo made of two glass sheets. Its dimensions are 800 mm high and 200 mm wide. Between the two panes a thin frame of stainless steel (1.1 mm thick) is sandwiched. This frame fixes the gap between the two glass sheets, and at the same time it contains the beads at the sides (see Fig.~\ref{fig:sistexp}). At the bottom, there is an orifice, its size being settled at 15.9 mm. Although this length is somewhat arbitrary, it is a compromise. On the one hand, the opening is well above the limit for jamming \cite{Janda1,To2,Zuriguel}, and therefore we have never observed an obstruction of the exit. On the other hand, the experiment is less demanding if the flow velocity is not too high, and as the outflow depends on the size of the opening, it cannot be arbitrarily large.

The granular sample consists of monodisperse stainless steel beads with a diameter of $1.00\pm0.01$ mm. The grains are poured along the whole width of the silo from a hopper at the top. The granular deposit within the silo consists of a monolayer of particles, because the gap between the glass sheets is slightly larger than the beads. Then a small overlap between the spheres in the images is possible. In our device, however, the overlaps detected are smaller than about $2~\%$ of the particle diameter. The overlap is likely to be even smaller if one considers that flowing beads are not necessarily in contact. Finite size effects in the lateral dimensions (corresponding to the width of the recipient) can be neglected as the silo dimensions (width and height) are much bigger than the size of the beads and the width of the opening at the bottom. The same set-up was used in previous works \cite{Mankoc,Janda1}; those previous results were taken as benchmarks for some of the data provided in this article.

Recordings were obtained with a high speed camera (Photron model FASTCAM 1024 PCI). With suitable optics and lighting, a small zone of the silo can be video taped with high resolution. We will now describe these zones where measurements have been performed, displayed in Fig.~\ref{fig:sistexp}. Four heights were chosen, viz. $y=30$, $65$, $155$ and $355$ mm ($y$ is measured from the bottom of the silo). A region of interest was then defined as a rectangle of 256 x 1024 pixels (width by height), the bottom side of the rectangle placed at the above mentioned heights. The spatial resolution is of 50 pixels per mm, so the rectangle corresponds to about 5 x 20 mm. These four zones are centered in the vertical axis of the silo, and are marked with rectangular patches in Fig.~\ref{fig:sistexp}. Under convenient illumination, each bead reflects a bright spot, as seen in the same figure, where a single frame is shown as an example. Films of about 3 seconds were recorded of the four zones at a speed ranging from 2000 to 5400 frames per second (the recording speed is changed so as to optimize the displacement of the beads from frame to frame). The length of the films was limited by the memory available in the recording system.

In order to rule out the possibility that the observation zone somehow influenced the results, we also recorded a second set of films changing a little the region of interest. In this second set, the window was a square of 512 x 512 pixels, with the bottom side at the same height than the previous rectangles. The recording speed was also changed by a small amount. As no differences were perceived in the results, in this article we will just show data corresponding to one set of recordings.

Once the films were recorded and stored, we processed them with a particle tracking software that was written in our laboratory to deal with a similar situation \cite{progmp}. The procedure involves splicing the film into frames, detecting the spots reflected by the beads and linking up the positions of the beads in adjacent frames. As the beads move less than one radius from one frame to the next, this can be done unambiguously. The software can pinpoint the position of the beads with subpixel resolution; the accuracy of the procedure has been checked with stationary beads, yielding a figure of $\pm 0.05$ pixels. In this way, we have been able to obtain about one thousand bead tracks for each film (order of magnitude).

Another series of recordings was performed in order to obtain the vertical velocity $v$ along a horizontal line. With this aim, we recorded a set of films with much lower magnification (about 5 pixels per mm), of a thin horizontal band (1024 x 64 pixels), spanning almost the whole 200 mm of the layer and centered at the vertical axis. The colored horizontal lines shown in Fig.~\ref{fig:sistexp} indicate those regions. The speed of the recording was much smaller in these cases (500 or 1000 frames per second usually).

The procedure carried out in this case was the following. A series of short films (typically 24 recordings, each one lasting about 0.1 second) were registered. Then for each particle, the mean vertical velocity was calculated. This amounts to fitting a straight line to the tracks and calculating the slope. All the data were aggregated into a single file for each zone. About $10^5$ individual particles (order of magnitude) were detected and processed at each height. For completeness, the mean velocity field for the whole silo, albeit with much smaller resolution, obtained with a PIV technique, can be seen in \cite{Janda1}.

\section{Velocity profile along the horizontal dimension}

\label{sec:veloprofile}
\begin{figure}
\includegraphics[width=\columnwidth]{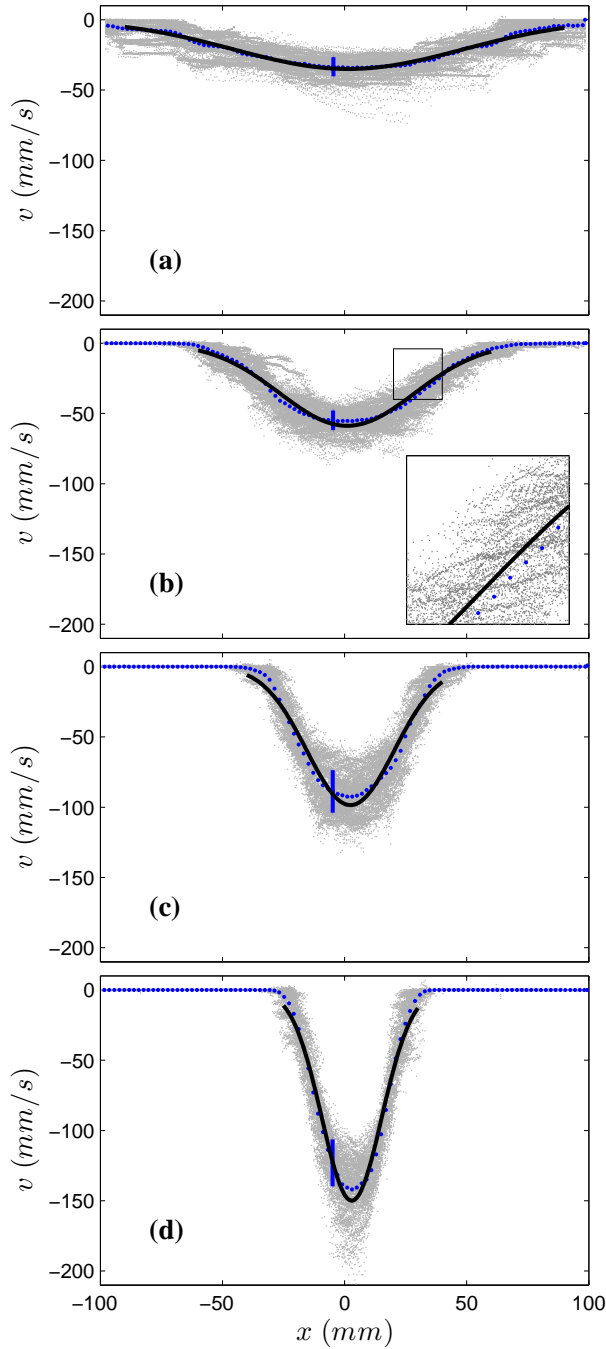}
\caption{
\label{fig:perfiles16}
(Color online) Vertical component of the velocity at different heights. The measurements were taken at the horizontal lines depicted on Fig.~\ref{fig:sistexp}: \textbf{a}, $y=355$; \textbf{b}, $y=155$; \textbf{c}, $y=65$; \textbf{d}, $y=30$. The cloud of small gray points corresponds to the velocities of individual particles. Blue dots correspond to the local mean (obtained every 2 mm approx.) A bar is provided displaying the standard deviation of the measurements around the mean (to avoid cluttering the figure just one bar is shown at $x\approx -5$ mm). Black solid lines correspond to the fit of Eq.~\ref{eq:vy}. Note that all the axes are at the same scale. The inset in plot \textbf{b} is a zoom of the zone indicated with a frame.
}
\end{figure}

As explained in the former section we have carried out the measurements of $v$, spanning all the width of the silo, at four heights. When preparing an experimental realization, it can be observed with the naked eye that particles tend to group in large clusters with preferred orientation order; the frontiers between them are clearly visible. It is not surprising, then, that the particles move orderly, displaying groups of data points with similar velocities (Fig.~\ref{fig:perfiles16}). In order to obtain suitable statistical averages, close to $10^5$ particle velocities were obtained at each height. Moreover, these measurements are the accumulation of more than 20 different runs, taken at different times. It is only after performing such a large amount of measurements that averages fit neatly to Eq.~\ref{eq:vy}. In Fig.~\ref{fig:perfiles16} we have plotted the average velocities along with the Gaussian fit.

\begin{table}
\caption{
\label{tab:tableBQ}
Values of the parameters $B$ and $Q$ (Eq.~\ref{eq:vy})
as given by the fit shown in Fig.~\ref{fig:perfiles16}.}
\begin{ruledtabular}
\begin{tabular}{lll}
$y$ (mm)&$B$ (mm)&$Q$ (beads per second)\\
\hline
355 & 3.07 & 4100\\
155 & 2.46 & 4100\\
65 & 2.46 & 4400\\
30 & 2.50 & 4600\\
\end{tabular}
\end{ruledtabular}
\end{table}

Although Eq.~\ref{eq:vy} is strictly valid only for a \emph{point-like} orifice at the bottom, we used it to fit the experimental observations leaving two free parameters: $B$ and $Q$; the values obtained are shown on Table \ref{tab:tableBQ}. The values of $Q$ are in reasonable agreement with the measured flow rate ($Q=4600$ beads per second). An alternative procedure could have been to fix $Q$ and leave $B$ as  the only free parameter (the results are similar). In any case, it is noteworthy that the values of $B$ are close to 2, which was the initial guess of Nedderman and other authors \cite{Nedderman}. We do not observe a neat dependence of $B$ with height except for the upper region; but this could be due to other factors: the velocities are much smaller, and the region where beads move downwards reach the lateral walls, so the boundaries could influence the velocity field.

As remarked, the agreement between the experimental data and Eq.~\ref{eq:vy} is good \emph{for the average velocities}. However, the deviations of the velocities of individual particles from the average are quite big, as revealed by the standard deviation. Moreover, collective motion of grains can be discerned if one looks closely at the data points, as shown in the inset in Fig.~\ref{fig:perfiles16} \textbf{b}, where aligned clouds of points show these events. This means that the motion of particles is not just noisy, but also that the grains move in \emph{clusters}. It is therefore pertinent to take a closer look on the movement of individual grains.

\section{Velocities of individual particles}
\label{sec:vi}

\begin{figure*}
{\includegraphics[width=2\columnwidth]{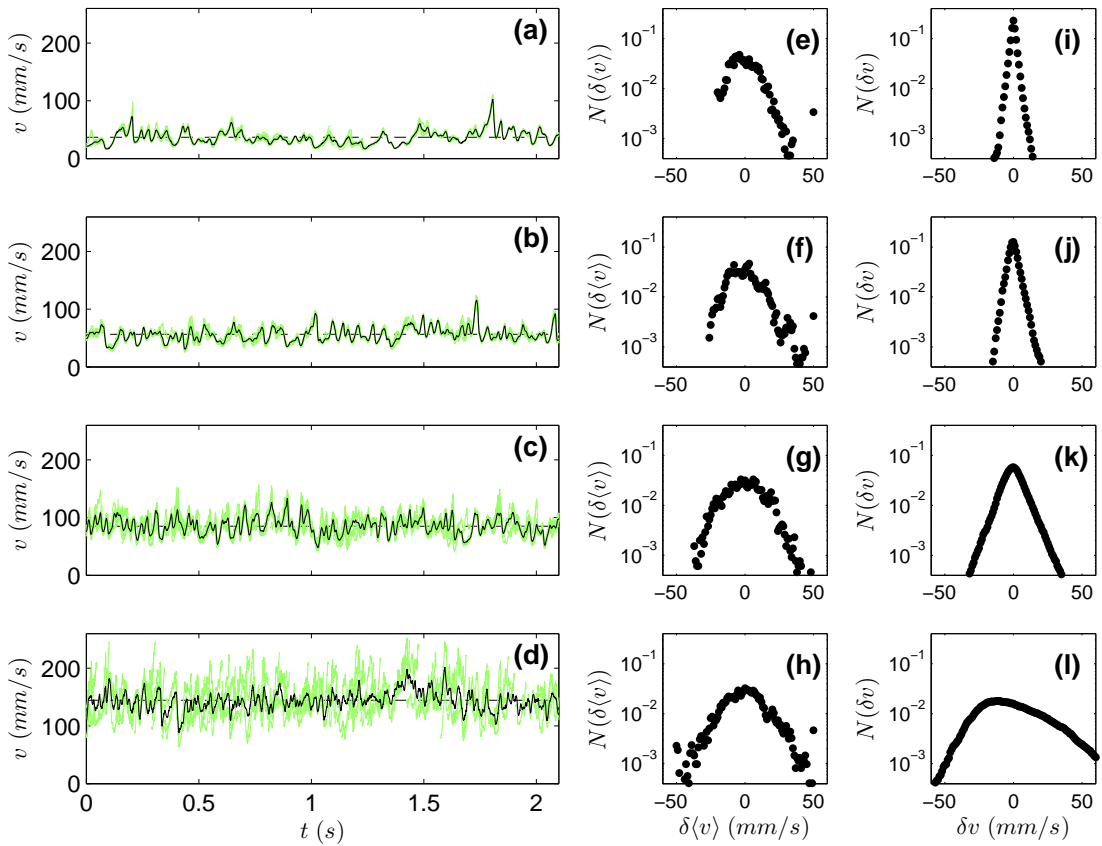}}
\caption{
\label{fig:fgv}
(Color online) On the left column (plots \textbf{a} to \textbf{d}), the vertical component of the velocity of individual particles $v$, obtained as explained in the text, is shown with green (gray) lines. The four plots correspond to the rectangular regions depicted in Fig.~\ref{fig:sistexp}: \textbf{a}, $y=355$; \textbf{b}, $y=155$; \textbf{c}, $y=65$; \textbf{d}, $y=30$. Only 5\% of the traces are represented to avoid cluttering the graphics. The solid black lines are the instantaneous averages $\langle v \rangle$, taking in this case all the measured particles, not just the ones represented here. The dashed horizontal line corresponds to the mean vertical velocity $V$, averaged over all the time for all the beads. Note that all the axes are at the same scale to allow easy comparison.\\
The plots on the center column (plots \textbf{e} to \textbf{h}) display the normalized histograms of the fluctuations of $\langle v \rangle$ (the average velocities represented in the left column with a solid black line) around the mean, $V$. Each plot corresponds to the same zone than the one to its left. Note the logarithmic scale on the vertical axis. All the axes are at the same scale to allow easy comparison.\\
The right column (plots \textbf{i} to \textbf{l}) corresponds to the normalized histograms of the fluctuations of the individual velocities $v$ around the average $\langle v \rangle$. Each plot corresponds to the same zone than the one to its left. Note the logarithmic scale on the vertical axis. All the axes are at the same scale, which is also the same as in the center column, to allow easy comparison.}
\end{figure*}

We will now focus our attention on the path of individual particles resolved in time, and we will describe it through the instantaneous velocity of each particle. As explained in Sec.~\ref{sec:setup}, the frame rate of the recording was tuned to obtain a point at a sampling frequency such that the displacement of the particles from frame to frame is close to the spatial resolution of the system. As the velocity implies a derivation, if one performs a straightforward calculation the noise is too large. In order to avoid it, we have calculated the velocity averaged at a length scale corresponding to a bead radius. The method is the following. The track of one particle is taken, and from each point of the track, the location in the trace at a distance equal to the bead radius is searched; the time elapsed from the current point to that position is calculated by interpolation; and from this time lapse the velocity was computed as the radius divided by the elapsed time. This figure yields the speed (velocity modulus), but as the direction of the bead movement can also be obtained, we next extract the vertical component $v$. In the following we will deal exclusively with the absolute value this component. (Remark, however, that the horizontal component of the velocity is much smaller than the vertical one, so the results we will present do not change noticeably if the speed is considered instead of $v$). In fact, the method explained amounts to low-pass-filtering the vertical velocity at a length scale of a bead radius. This procedure was carried out with more than 1000 beads for each one of the four rectangular patches depicted in Fig.~\ref{fig:sistexp}. The results are shown in Fig.~\ref{fig:fgv}.

The average vertical velocity of all the particles in the observation region at a given instant, that we will call $\langle v \rangle$, can be calculated from the velocities of individual particles. It is also represented in the left column graphics of Fig.~\ref{fig:fgv} with solid black lines. Then the mean vertical velocity, $V$, averaged over time, is also obtained; it is represented in the same plots with a dashed horizontal line. The values obtained in the four observation zones, from top to bottom, are $V=$ 37.0, 56.6, 85.1 and 144.4 mm/s.

Direct observation of the leftmost column of Fig.~\ref{fig:fgv} indicates that the fluctuations of the average speed $\langle v \rangle$ increase as the orifice is approached. But not only that; the deviations of individual particles from the average also seem to be enhanced. It can be appreciated with the naked eye that at the top of the silo particles move in concert, as a group, closely assembled around the mean; at the bottom, the dispersion around the average seems more important. This inspired us to separate the fluctuations into two: the variation of the average itself, and the deviations of individual particles from the average. This is not a standard procedure (for instance, when studying turbulent fluids), because in usual situations there is a well defined time averaged velocity $V$ that does not fluctuate wildly; in our case this is not so. In the following section we will take a closer look into it these fluctuations.

\section{Fluctuations}
\label{sec:fluctuations}

The fluctuations of the velocities are thus separated into two categories: the fluctuations of the average velocity $\langle v \rangle$ around $V$, and the fluctuations of individual velocities around the average $\langle v \rangle$. The former can be regarded as the fluctuations of a certain advective velocity field (due to the gravity), while the second could be assigned to the random movement of the particles around the advective field.

\begin{figure}
{\includegraphics[width=0.9\columnwidth]{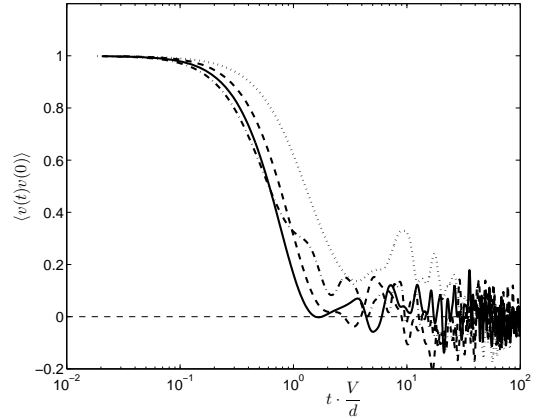}}
\caption{
\label{fig:acvrad}
Autocorrelation of the average velocity $\langle v \rangle$. The horizontal axis has been rescaled by the average time it takes for a bead to fall its own diameter. The four lines correspond to the four heights where measurements were performed: dot-dashed line, $y=355$; solid, $y=155$; dashed, $y=65$, and dotted, $y=30$.}
\end{figure}

The fluctuations of $\langle v \rangle$ are shown on the center column of Fig.~\ref{fig:fgv}. Although they are not exactly Gaussian, one can discern a shape resembling a parabola (in semilogarithmic scale). This probably means that the average velocity is just noisy and that the fluctuations arise from a stochastic process. In fact, the autocorrelation function of $\langle v \rangle$ (Fig.~\ref{fig:acvrad}) shows that it is uncorrelated for times longer than the delay that it takes for a bead to fall its own diameter. Similar behavior was found in numerical simulations \cite{Tewari}. An important remark concerning this should be noted: the fact that the autocorrelation of $\langle v \rangle$ goes to one for short times was something to be expected from the method we used to calculate the velocities, which erases out the fluctuations at times smaller than $t \cdot \frac{V}{d}$. The relevant result here is that the autocorrelation decays to zero for times larger than this. In Fig.~\ref{fig:acvrad} the rescaled time lag is shown to stress the point that at about $t = \frac{V}{d}$ the average velocity $\langle v \rangle$ quickly becomes uncorrelated. In such a representation, the autocorrelation decays later at the bottom of the silo. But let us note that in fact the \emph{unrescaled} decay times are larger where the velocity is smaller (\emph{i. e}, at the top of the silo, the autocorrelation of $\langle v \rangle$ decays later than at the bottom).

On the other hand, the fluctuations of $v$ around $\langle v \rangle$ are clearly not Gaussian (right column of Fig.~\ref{fig:fgv}). They are pointed and sharp, meaning that the deviations from the average are \emph{smaller} than if they were the result of a random process. This is a hint of a collective, correlated motion of the particles. This feature is much more remarkable far from the exit orifice, at the top of the silo, where the mean velocity $V$ is small. At the bottom, the distribution is wider and less sharp. This again agrees with previous numerical simulations \cite{Arevalo}.

The comparison of both sets of fluctuations (the deviations of $\langle v \rangle$ around $V$ and the deviations of the individual velocities $v$ around $\langle v \rangle$) reveals that their amplitudes are of the same order of magnitude (Fig.~\ref{fig:fgv}). Another common feature to both sets of fluctuations, which is in good agreement with previous observations, is that the distributions are slightly slant to the right. That is, there are slightly more large deviations towards the region of higher velocities (meaning falling more rapidly) than towards the region of smaller velocities. As mentioned above, the main difference between the PDFs is that while the deviations of $\langle v \rangle$ around $V$ are approximately Gaussian, the deviations of $v$ around $\langle v \rangle$ display a pointed shape revealing collective motion. This suggest an underlying process of anomalous diffusion.

\begin{figure}
{\includegraphics[width=0.9\columnwidth]{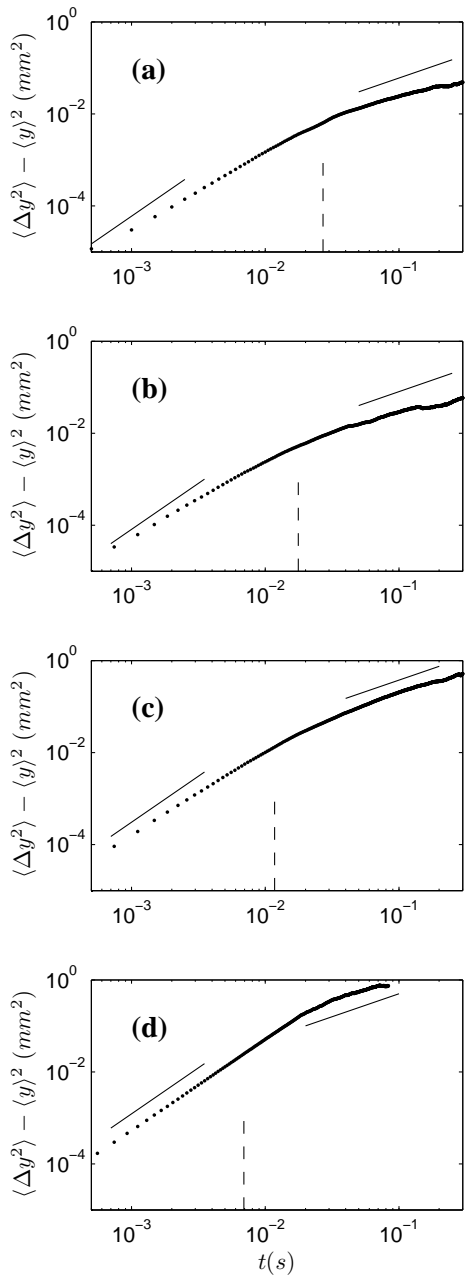}}
\caption{
\label{fig:fgrafvar}
The variance of the displacements of the beads, calculated as explained in the text,  at the four heights \textbf{a}, $y=355$; \textbf{b}, $y=155$; \textbf{c}, $y=65$; \textbf{d}, $y=30$. Note that all the axis are at the same scale. The solid lines are a guide for the eye: in each plot the line at the left as a slope of 2 and the one at the right a slope of 1. The dashed vertical lines indicate the instant at which the particle has covered, in average, a distance equal to its own diameter.}
\end{figure}

In order to investigate this, we have computed the variance of the displacements of the beads. In this calculation the ``mean path'' must be subtracted from the track of each particle. Calculating directly the mean position of a group of beads can be tricky because particles entering or exiting the observation window must be explicitly taken into account. Instead, we have integrated the average velocity $\langle v \rangle$ obtained previously to calculate the mean displacement. This quantity at each instant is then subtracted from the particle tracks to compute the variance of the vertical displacements  $\sigma^{2}_{y}=\langle (y - \langle y \rangle )^2 \rangle$. The same method can also be performed for the $x$ component, but as the results do not add new information we will only present the variance for the $y$ component (Fig.~\ref{fig:fgrafvar}). In particular, an explicit comparison is drawn with a diffusive regime ($\sigma^{2}_{y} \sim t $) and with a ballistic regime ($\sigma^{2}_{y} \sim t^2$). It has been pointed out, with the support of experimental data \cite{Harich}, that fluctuations of dense grains should generically display a ballistic regime for small times; then, as time increases, a regime might develop showing the cage effect \cite{Reis}, in which basically the fluctuations remain at the same level with time (\emph{i. e.} $\sim t^0$); finally, a diffusive regime is expected for long enough times. Times can be gauged in terms of the distance traveled by the particle: times smaller than the interval it takes for a bead to move a distance equal to its own diameter are small; the cage effect, if it exists, and the transition to a diffusive regime, should be more conspicuous from there on. In Fig.~\ref{fig:fgrafvar} we have marked the moment at which the particle has traveled a distance equal to its diameter (in average).

The picture that emerges is similar to that of Brownian motion resolved for very small times \cite{Li}, except that an intermediate stage occurs where the particle is ``trapped in a cage'' by its neighbors. Our data are not able to display explicitly this scenario, because the time that one can continuously follow a particle is limited by the observation window -- or else by the gradients inside the observation window, if one tries to consider bigger and bigger windows. We can nevertheless identify a region for small $t$ where the fluctuations scale approximately as $t^2$. This scaling evolves in time and for displacements larger than a diameter the fluctuations appear to scale with time subdiffusively, \emph{i. e.} the slope is smaller than 1 (Fig.~\ref{fig:fgrafvar}). Remarkably, this evolution is particularly evident for the region near the orifice (Fig.~\ref{fig:fgrafvar}.d) where the ballistic regime extends beyond the limit of one particle diameter. In this region the particles seem to be in a less dense state, where displacements are manly determined by the gravity field, that accelerates the particles and causes the collisions -- which is the mechanism governing the diffusive motion.

\section{Discussion}
\label{sec:conclusions}

We provided experimental data about the microscopic movement of particles in a 2D dense granular flow. The spatial and time resolution allowed us to explore the collective motion. It has been show that the averaged vertical velocity profile can be fitted to a Gaussian function which is solution of a the diffusive-like equation. Although the instantaneous velocity of the particles display correlated motions, long time averages show the typical velocity profile reported in different experimental situations \cite{choi,moka,kumaran}. It has been sometimes argued that systems consisting of monodisperse particles can not be used to study granular flows due to its tendency to crystallize. Nevertheless these results show that even in this case the diffusive approach can be used to fit the average velocity profile. Importantly, the election of monodisperse beads in a 2D configuration gives the opportunity of test the fundamental assumption of the theoretical models related with the existence of a well defined temporal scale of the particle displacement at microscopic level. As we show in Sec.~\ref{sec:fluctuations} the particle motion can be split into a global or \emph{advective} movement and another one ``on top'' of it where the diffusive approach can be considered. The former displays a commonplace Gaussian PDF of the velocity fluctuations, while the latter shows a strong non-Gaussian PDF, conspicuously indicating collective motion. The Gaussian-like fluctuations seem to indicate the randomness of the global deformation of the granular layer. On the other hand the evolution of the autocorrelation function of the averaged velocity $\langle v \rangle$ in the vertical direction (Fig~\ref{fig:acvrad}) shows that the correlation of this variable becomes negligible at the time scale of one particle displacement, regardless of the mean velocity $V$. Finally, the variance of the path of individual particles with respect to the mean path, displays a ballistic regime for small displacements --suggesting that the global dynamics is governed by gravity at these scales. The anomalous regimes reported by other authors \cite{choi} could presumably be related to the combination of individual and collective motion. In addition, it seems that the variance at long time scales are anomalous, which could be related to intrinsic features of the advective field.

In summary, the study of the deviations of individual velocities $v$ from the instantaneous average $\langle v \rangle$ has revealed inspiring in order to clarify the origin of the microscopic dynamics in a silo discharge process.

\begin{acknowledgments}
We thank J.M. Pastor for his help. This work has been financially
supported by Projects FIS2008-06034-C02-01 (Spanish Government),
and PIUNA (Universidad de Navarra). A. J. thanks Fundaci\'on Ram\'on Areces and Asociaci\'on de Amigos de la Universidad de Navarra for a scholarship.
\end{acknowledgments}

\end{document}